\documentclass[11pt,twocolumn]{article}

\usepackage[utf8]{inputenc}
\usepackage[T1]{fontenc}
\usepackage{geometry}
\usepackage{amsmath}
\usepackage{amssymb}
\usepackage{graphicx}
\DeclareGraphicsExtensions{.pdf,.png,.jpg,.jpeg}
\usepackage[numbers,sort&compress]{natbib}
\usepackage{url}
\usepackage{hyperref}
\usepackage{color}
\usepackage{soul}
\usepackage{setspace}

\hypersetup{
    colorlinks=true,
    linkcolor=blue,
    citecolor=blue,
    urlcolor=blue
}

\title{From the perspective of perceptual speech quality: the robustness of frequency bands to noise}

\author{
    Junyi Fan \\
    Department of Computer Science and Engineering \\
    The Ohio State University, Columbus, Ohio 43210, USA \\
    \and
    Donald S. Williamson \\
    Department of Computer Science and Engineering \\
    The Ohio State University, Columbus, Ohio 43210, USA
}

\date{\today}

\begin{document}

\maketitle

\begin{abstract}
Speech quality is one of the main foci of speech-related research, where it is frequently studied with speech intelligibility, another essential measurement. Band-level perceptual speech intelligibility, however, has been studied frequently, whereas speech quality has not been thoroughly analyzed. In this paper, a MUltiple Stimuli with Hidden Reference and Anchor (MUSHRA) inspired approach was proposed to study the individual robustness of frequency bands to noise with perceptual speech quality as the measure. Speech signals were filtered into thirty-two frequency bands with compromising real-world noise employed at different signal-to-noise ratios. Robustness to noise indices of individual frequency bands were calculated based on the human-rated perceptual quality scores assigned to the reconstructed noisy speech signals. Trends in the results suggest the mid-frequency region appeared less robust to noise in terms of perceptual speech quality. These findings suggest future research aiming at improving speech quality should pay more attention to the mid-frequency region of the speech signals accordingly.
\end{abstract}

\footnotetext{
Copyright \textcopyright{} 2024 Acoustical Society of America.
This article may be downloaded for personal use only.
Any other use requires prior permission of the author and
the Acoustical Society of America.

The following article appeared in
J. Acoust. Soc. Am. 155, 1916--1927 (2024) and may be found at
https://doi.org/10.1121/10.0025272
}

\section{Introduction}
\label{sec:1}

Speech quality and intelligibility are both vital perceptual measurements of speech signals and are frequently used to assess the performance of speech-processing algorithms, including those for speech enhancement \cite{loizou2007speech, paliwal2011importance}, speech separation \cite{hershey2016deep, 8369155} and text-to-speech (TTS) synthesis \cite{shen2018natural}, to name a few. Clean and anechoic speech signals are expected to have high levels of quality and intelligibility, while signals that have been corrupted by additive noise, reverberation, or further processing may degrade in quality and intelligibility \cite{loizou2007speech}. Speech quality, as an intricate psychoacoustic phenomenon, is a highly subjective metric. It is difficult to precisely define since it involves multiple perceptual dimensions such as naturalness and listening effort \cite{Grancharov2008}. Intelligibility, on the other hand, is more objective, where it is defined and measured by the percentage of speech elements that are correctly recognized by the listeners. Due to the vagueness of its definition and its intricate nature, speech quality is not as well understood as its counterpart speech intelligibility.

Efforts to investigate methods for assessing speech quality have been made in the past, although the outcomes of these studies may have not been well utilized to extensively examine speech quality from a psychoacoustic perspective. Methods of subject listening tests were proposed by \citet{1988}, which can be broadly summarized as methods based on either relative preference or quality ratings. Many subsequent listening tests were designed based on these two concepts, such as MUSHRA \cite{mushra} which assesses speech quality by assigning quality rating scores. Objective methods for evaluating speech quality were also studied in the hope of avoiding expensive and time-consuming subjective tests. One objective method known as perceptual evaluation of speech quality (PESQ) by \citet{pesq} has been frequently deployed to assess speech quality as it is able to provide quality scores that are highly correlated to those from subjective tests \cite{pesqcompare}. A review regarding speech quality assessment was also presented by \citet{Loizou2011} where subjective and objective speech quality assessment methods were discussed more thoroughly. In recent years, thanks to the rapid development of deep learning, models based on deep neural networks \cite{Dong2020TowardsRO, kareddy2020dnsmos} were implemented to better assess the speech quality by overcoming the weaknesses of the traditional methods. It is believed that the outcomes of the previous works have undoubtedly benefited various areas that aspire to improve perceptual speech quality. Nevertheless, with the same potential that these findings can also benefit research that focuses on understanding perceptual speech quality itself, these research problems have not been paid equivalently due attention.

Such a lack of thorough examinations of perceptual speech quality frequently brings out inevitable problems in research areas involving speech signals. In speech enhancement, for instance, a good amount of research treats speech quality and intelligibility without too much fine discrimination while designing the speech enhancement models, although no necessary connections between them were clearly observed in previous studies \cite{loizou2007speech}. As a result, such approaches may produce speech enhancement algorithms that manage to improve one while failing to improve the other \cite{5428850, wang2008time, 1415651, kim2009algorithm, healy2013algorithm}. Understanding the inherent similarities and differences in speech quality and intelligibility is significant to speech perception, as well as areas that may benefit from it. With such information, it becomes possible for researchers to incorporate these overlooked ideas in the early stage of algorithm design, which may further fine-tune the performance of the related research outcomes. Furthermore, it is desired that these techniques be deployed in many real-world applications that can bring positive impacts to those including the communication industry, individuals with hearing impairments, and speech enhancement researchers, to name a few.

It is widely understood that different acoustic features can be commonly observed across speech frequency bands. Such a phenomenon further leads to different phonetic and linguistic behaviors in each specific region. These different behaviors can potentially result in the non-constant noise robustness of the individual frequency bands. Work by \citet{PMID:11931319} in analyzing the relative importance to speech intelligibility of different intensities provided a series of intensity-importance functions. The functions, however, did not remain unchanged within the speech dynamic range at different frequencies, which suggests an inconsistency in the behaviors of different speech frequency bands. \citet{apouxbacon} also concluded the importance of four speech bands, investigated in their study for identifying consonants in quiet and noisy environments, appears to be different, revealing that noise might potentially have different levels of influence on speech intelligibility across frequency bands. More evidence has been shown in similar studies regarding speech band importance \cite{kasturi, apoux} that different frequency bands in noisy speech may contribute differently to the overall speech intelligibility. Although it is unfair to draw the conclusion that these bands are not consistently robust to noise without systematic examinations but simply from such previous observations, or even to suggest the two are correlated, it potentially introduces the idea that the robustness of frequency bands to noise may present similar inconsistencies across the spectrum due to similar underlying impacts. 

The recent work closely related to the topic by \citet{sarah} studied the noise susceptibility across various critical speech bands with speech intelligibility as the measurement. As indicated by the previous observations, noise susceptibility in terms of speech intelligibility was proven to greatly vary across the speech spectrum. A basic pattern of noise susceptibility was observed across the frequency spectrum, where in general the lower half of the spectrum appeared to be less susceptible to noise. However, conclusions drawn from the experiments examining speech intelligibility cannot be simply extended to speech quality due to the inherent discrepancies in their nature. Moreover, artificial noise was deployed in the experiment which might impact the extensiveness of the conclusions. Randomly selected frequency bands with the remaining bands missing in trials may require a larger scale of experiments to eliminate the randomness and might also bring unknown impacts to the results. Due to the same reason, the target signals were also no longer broadband and it is unclear how this would potentially limit the generalization of the findings in the study.

Nevertheless, few studies have investigated such a hypothesis with speech quality as the measurement, where these prior studies aim to understand perceptual speech quality with slightly different foci. In~\citet{hansenoriginal}, the importance of frequency bands based on pairwise preference tests was examined, resulting in band-specific detection thresholds for distinguishing between pairs of signals. The speech frequency bands were found to be appropriate for designing an objective quality measure. However, the experiments evaluated band-specific quality based on signals after speech transmission (e.g., telephony speech), and importance was not assessed based on real-world noise since modulated white noise was only considered. Moreover, the problem of how phase information may impact speech quality has been studied by different groups of researchers \cite{1163920, paliwal2011importance, 7038277, zhang2020investigation} in the hope of incorporating the outcomes into the related techniques and applications to improve their performance. These studies, however, inspected speech quality from a very different perspective of phase information, which suggests little to the hypothesis proposed in this study. Works to study the non-intrusive speech quality assessment methods conducted extensive scales of subjective listening tests and provided a considerable amount of quality rating scores \cite{Dong2020TowardsRO, kareddy2020dnsmos, yi2022conferencingspeech}. The correlations between the subjective results obtained from the tests and objective results obtained from their proposed methods were analyzed, with many of them presenting outstanding performance. However, these studies merely collected quality rating scores through listening tests mainly as data for their proposed models instead of trying to better understand speech quality itself. Most importantly, prior work did not examine band-level noise robustness, which exhibits great potential to facilitate both people's understanding of the perceptual speech quality and to improve the performance of various techniques addressing the issue of poor speech quality in speech processing.  

Previous studies suggest that the potential discrepancies in the robustness of various frequency bands to noise should be investigated from the perspective of speech quality. To this end, we proposed an approach, inspired by the MUltiple Stimuli with Hidden Reference and Anchor (MUSHRA), to collect perceptual speech quality responses from human subjects recruited on an online crowdsourcing service platform.  
The speech signals studied in the experiment were constructed from real-world clean speech stimuli and noise recordings. Speech and noise materials were filtered into frequency bands and then added together at various signal-to-noise ratios (SNRs). Participants were then instructed to assign quality ratings to the speech signals during the MUSHRA-inspired listening tests. 
MUSHRA has been widely deployed to evaluate the perceptual quality of speech signals and is proven to be capable of rendering statistically significant results by numerous past works \cite{sporer2009statistics}. Although it is well-known for its efficiency in its implementation and the process of response collection, the scale of the experiment in this study is still inevitably extensive due to the requirement of the band-level examination. Empowered by the online crowdsourcing service, a large number of subjects can be recruited to participate in the listening study \cite{Schoeffler2014TowardsTN, 7471749}. Online MUSHRA-based listening tests have been shown to provide no significantly different results compared to those collected in controlled environments with selected participants, professional audio equipment, and the same experiment setups \cite{nodifference}. The experiment was deployed based on materials recorded in real environments as suggested by recent works \cite{McLaren2016THESI, reddy2020interspeech}, unlike some of the previous studies where artificial noise was added to the background \cite{apoux, reverb}. The more realistic materials are expected to produce more practical and convincing conclusions to better benefit the real-world application situations, as it is expected to be able to capture more intricate details of the real environments. Band-level robustness to noise indices were calculated based on the collected quality rating scores and were further compared to the results obtained from the speech intelligibility study to investigate the potential correlations.

The paper is organized as follows. A detailed methodology is presented in Section~\ref{sec:2}. Results are presented in Section~\ref{sec:3}. Discussions and conclusions are presented in Section~\ref{sec:4} and Section~\ref{sec:5}, respectively.

\section{Method}
\label{sec:2}

\subsection{Subjects}
\label{subsec:2:1}

The subjects in the crowdsourced subjective listening tests were all recruited from Amazon Mechanical Turk (MTurk) \cite{paolacci2010running}. In the preliminary listening test, responses from 275 normal-hearing subjects were qualified to be analyzed. The ages range from 20 to 74, with a mean of 38.5. Among all the subjects, 165 were reported to be male and 109 were reported to be female. In the primary listening test, responses from 165 normal-hearing subjects were qualified, with their ages ranging from 23 to 67 with a mean of 37.4. In this group, 90 of the subjects reported to be male, and 75 of them reported to be female. All subjects were native speakers of American English and physically living in the US, with self-reported normal hearing capacity and high MTurk approval ratings. They also indicated that they were physically in a quiet environment and using listening devices that could help eliminate the surrounding environment noise while participating in this study on a computer. The subjects were allowed to take breaks or withdraw at any point during the test if needed. The study was approved by the Ohio State University's Office of Responsible Research Practices before being published. A monetary incentive was provided for all subjects submitting qualified responses.

\subsection{Speech stimuli and processing}
\label{subsec:2:2}

Both the speech and noise materials in this study were obtained from the datasets provided by the 3rd CHiME Speech Separation and Recognition Challenge \cite{chime3}. The clean speech stimuli provided by the corpora were recordings of sentences from the WSJ0 corpus \cite{wsj0} spoken in quiet real environments with a moderate speaking rate. The talkers include both males and females and have a general American English accent. The noise signals were recorded in 4 real-world locations, including a pedestrian area (PED), cafe (CAF), public transport (BUS), and street junction (STR). STR mainly contains noise caused by passing traffic, while PED contains a great degree of close-range pedestrian noise. CAF contains noise greatly contributed by background competing speech, clicking noise of tableware, and occasional background music. BUS primarily contains vehicle engine noise and background speech. All of them also contain different levels of miscellaneous noise from the recording environments. For the preliminary experiment, 51 clean speech stimuli were indiscriminately drawn from the datasets, while for the primary experiment, a subset of 20 speech stimuli were drawn from the 51 speech stimuli used in the preliminary experiment. The noise signals were randomly drawn from all four aforementioned noise categories. Most speech stimuli presented in the experiment roughly range from 5 seconds to 15 seconds in length. The noise signals were shortened to match the exact length of their corresponding speech stimuli. Among the 51 individual noise clips included in the preliminary experiment, 13, 15, 9, and 14 were randomly chosen from PED, CAF, BUS, and STR, respectively. In the primary experiment, 5 noise clips from each noise category, which formed a subset of 20 noise clips in total, were further randomly selected from those used in the preliminary experiment, to ensure the balance of the number of noise clips from each category.
Both speech stimuli and noise signals were 32-bit audio files sampled at 16 kHz. These selected recordings consist of highly-varied audio situations in terms of the talkers' genders, the content of the speech materials, and the types of background noise, to name a few.

\begin{figure}
\centering
\includegraphics[width=0.75\columnwidth]{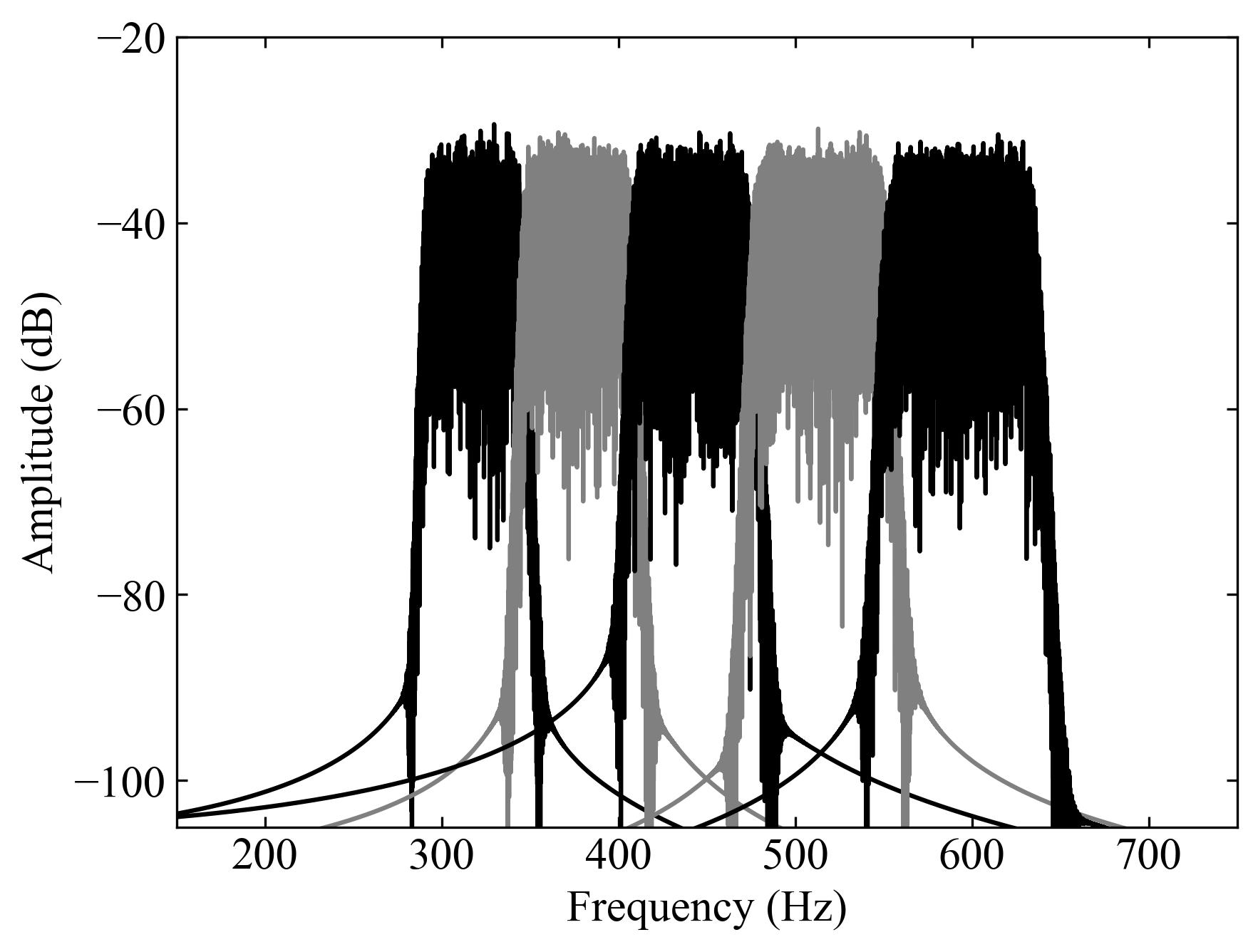}
\caption{A depiction of the frequency responses to extract bands 5 through 9 given a 90-second white-noise input signal.}
\label{bandfreq}
\end{figure}

Speech and noise materials were filtered into 32 contiguous frequency bands with center frequencies ranging from 100 to 7500 Hz, similar to what was done in \citet{apoux} with minor adjustments adopted to ensure the suitability for the current study. Details regarding the values of these center frequencies can be found in the later sections. Two cascaded 28th-order digital Butterworth filters were used for filtering. The standard zero phase digital filtering technique was incorporated to filter the input signals in forward and backward directions in order to avoid phase distortion in the output signals. Each individual band was filtered to be one ERB$_N$ \cite{erbaa} wide to resemble the normal-hearing human auditory system. The level of overlap amongst the bands can be visualized in Fig.~\ref{bandfreq}, where it presents the responses of 5 consecutive filters applied to band 5--9 to a 90-second white noise signal. The spectra also confirm that the filters applied to these bands were capable of generating slopes that exceeded 6 dB/octave with all 32 filters following a similar pattern.

To avoid unknown randomness and ensure broadband speech signals, all 32 frequency bands were present in the final reconstructed audio signals prepared for the experiments. In this fashion, to ensure noticeable quality differences among the reconstructed audio signals, a sliding window of 16 consecutive bands was treated as target bands, and the remaining bands were considered as extra bands in the preliminary experiment. Such target bands ranged from bands 0--15 to bands 16--31, rendering 17 different combinations. Each speech band combination was subsequently added with its corresponding noise band combination (i.e.~the two combinations have the same band components) to achieve one of six desired SNRs: -15, -10, -5, 0, 5, and 10 dB. More specifically, let $S$ denote a speech signal and $N$ denote a noise signal, while $S_i$ and $N_i$ denote the signals centered at the $i^{th}$ frequency band, $i \in \{0,1,\cdots,31\}$, with $i=0$ being the band at the lowest center frequency and $i=31$ at the highest center frequency. To construct the target bands, $T_m$, where $m \in \{0-15, 1-16, \cdots,16-31\}$, the set of frequency band signals within $m$ were first added together forming the speech signal, $S_m$. The same operation was applied to the noise signal, producing $N_m$. $S_m$ and $N_m$ were then added together to generate $T_m$. The process can be denoted as
\begin{align} 
T_m = S_m + \alpha N_m,
\end{align} 
where $\alpha$ is the scaling factor to ensure $T_m$ possesses one of the desired SNRs. The non-target bands (i.e.~the remaining bands that are not considered target bands) were not generated from these band combinations $m$, but instead, each one of them was first constructed by adding the individual speech band and its corresponding noise band (i.e.~the two bands have the same band number) with an SNR of 0 dB. This can be denoted as
\begin{align} 
E_i = S_i + \beta N_i,
\end{align} 
where $\beta$ is the scaling factor to ensure $E_i$ has a 0 dB SNR. $T_m$ and $E_i$ were then added together to generate the final reconstructed audio signal $X$, denoted as
\begin{align} 
X = T_m + \sum_{i \notin m} E_i.
\end{align} 
The selected SNRs and the band configurations were determined through a series of experiments, to ensure all reconstructed signals were proper for the experiments with being reasonably noisy but not too clean or distorted.

The primary experiment was further developed based on the results of the preliminary experiment. In the primary experiment, only three of the original six SNRs at -10, 0, and 10 dB were selected for the target bands. Furthermore, to ensure that each individual frequency band could appear the same amount of times in the reconstructed audio signals, 15 additional target band combinations were added to the experiment by constructing the target band combinations in a circular fashion. In this case, 15 more combinations from band 17--0 to band 31--14 were included, rendering 32 various band combinations in total including the original 17 combinations.

For both the preliminary and primary experiments, the clean references and anchor signals with an overall SNR of -15 dB were also used for calibration. In contrast to the recommended MUSHRA where low-range and mid-range anchors were included, the anchors in this study possessed an overall SNR of -15 dB to serve a similar purpose of calibration. They were, however, more appropriate for this study as they allowed the test to be calibrated more conveniently, where they were expected to serve as the low bound of the expected qualities of all signals, whereas it was unclear where the recommended low-pass version anchors might fit among the test signals. The volumes of all the reconstructed signals were normalized to the same energy level for the listening tests.

\subsection{Procedure}
\label{subsec:2:3}

The perceptual listening tests were implemented on Qualtrics to investigate the suspected subtle differences in the noise robustness of frequency bands. The variables are two-dimensional in various SNR values and band combinations. For both experiments, in each trial, the subjects listened to a group of audio signals to compare before assessing their overall qualities by assigning each of them a quality rating score between 0 and 100, representing extremely bad and extremely good quality, respectively. Unlike the more traditional five-point MOS scale recommended by \citet{five_scale}, a 100-point scale was used given the assumption that the quality differences among frequency bands can be so subtle that the precision of a five-point scale may not be sufficient. This assumption is proven to be reasonable by the results in the later sections. The test audio signals in each trial were generated by the same speech and noise materials with the same band combination but at different SNR levels, together with the corresponding clean reference and anchor hidden among them. Unlike the traditional MUSHRA, a labeled open reference was not provided in each trial along with the hidden reference in the hope of reducing the time needed for the experiment. Previous experience suggests such a variation does not have significant impacts on the results \cite{dong2020pyramid} and that it can be deployed to assess various types of audio situations \cite{nolabel}. Subjects were required to finish rating all signals in the current trial before being allowed to proceed to the next trial. Subjects could not go back to the prior trials after starting the current ones. Each subject was expected to complete 17 and 20 trials in the preliminary and primary experiments, respectively, as well as an additional practice trial preceding the formal ones to achieve familiarization. The practice trial was identical to the formal ones but presented with different speech materials that would not be heard again thereafter. The duration of the listening test was approximately 18 minutes on average. Each individual trial was evaluated 5 times to avoid randomness and bias and to achieve statistically significant results. No time limit was enforced and subjects could take a break whenever needed during the test as long as the test could be finished within 5 days. The majority of the subjects were only allowed to participate once.

\subsection{Data cleaning and index calculation}
\label{subsec:2:4}

All responses were carefully examined before being accepted as qualified results. Strict qualification metrics were deployed to detect any malicious response \cite{mturkcleaning}. Responses from a crowdworker were rejected if the task was finished in an unreasonably short amount of time. Responses were also rejected if random scoring was detected. Responses composed of an exceeding amount of unreasonable rating scores caused by careless human error were also rejected. 

Before calculating the noise robustness indices, the mean perceptual quality scores for each frequency band at different SNR conditions were calculated in an inverse moving average fashion. To this end, arithmetic means of quality scores provided by all corresponding subjects were calculated to provide the mean quality scores $A(m, r, o)$ for a specific audio signal $o$ with band combination $m$ at a certain SNR $r$. Subsequently, the quality scores $Q(m, r)$ for a specific band combination $m$ at a specific SNR $r$ could be calculated by averaging across all test signals $o$ sharing the same band combination $m$ and SNR $r$. All frequency bands were processed under the same condition and combined in the same way to reconstruct the signals for the experiment. It is assumed that given these conditions and the specific design of the experiments, all frequency bands contribute consistently to the perceptual qualities of the reconstructed audio signals. This assumption is similar to the premise on which some previous studies are based that speech intelligibility is modeled as total contributions of independent speech frequency bands \cite{ansi1997, apoux}. With such an assumption, given a certain SNR $r$, the individual quality score $B(i, r)$ for each band $i$, $i \in \{0,1,\cdots,31\}$ can be calculated by averaging across a group of quality scores $Q(m, r)$, where all band combinations $m$ contain band $i$. 
This can be denoted as
\begin{align} 
B(i,r) = \frac{1}{T} \sum_{m} Q(m,r)\, \mathbf{1}_{\{ i \in m \}}
\end{align}
where $\mathbf{1}_{\{ i \in m \}}$ returns $1$ when band $i$ is contained in $m$, and it returns $0$ otherwise and $T$ is the number of times $\mathbf{1}_{\{ i \in m \}}$ returns $1$.

Finally, to obtain the individual noise robustness indices of various frequency bands $i$, quality scores $B(i, r)$ at a certain SNR $r$ dB were normalized to $B_{norm}(i, r)$ by min--max normalization so that the highest score $B_{norm}(i_h, r)$ among the 32 scores is always 1 and the lowest score $B_{norm}(i_l, r)$ is always 0. The index score of a certain band $i$ was calculated as the mean of all selected $B_{norm}(i, r)$ where $r$ represents the selected SNR conditions to generate the final quality index scores.

\section{Results}
\label{sec:3}

\subsection{Preliminary experiment: perceptual quality scores by SNR}

\begin{figure}
\centering
\includegraphics[width=0.75\columnwidth]{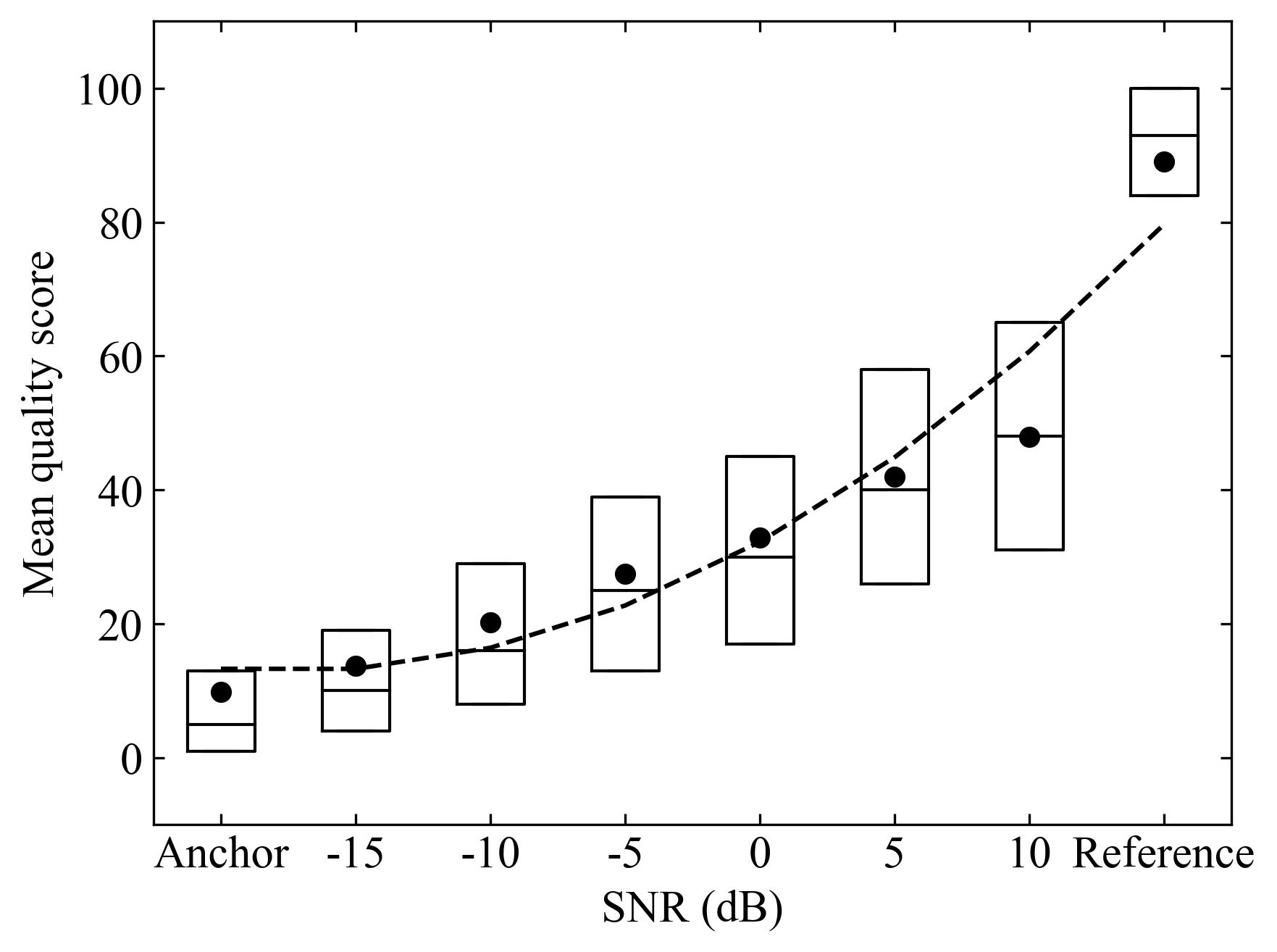}
\caption{Mean perceptual speech quality scores (e.g., circle markers) based on the listening test as a function of the SNR of the target bands in the preliminary experiment. Scores are calculated by averaging across all 17 band combinations. A second-order regression fit is present as the dashed line. Boxplots are also shown to indicate the 25th, 50th (median), and 75th percentiles at each SNR condition.}
\label{overall}
\end{figure}

\begin{figure*}
\centering
\includegraphics[width=\textwidth]{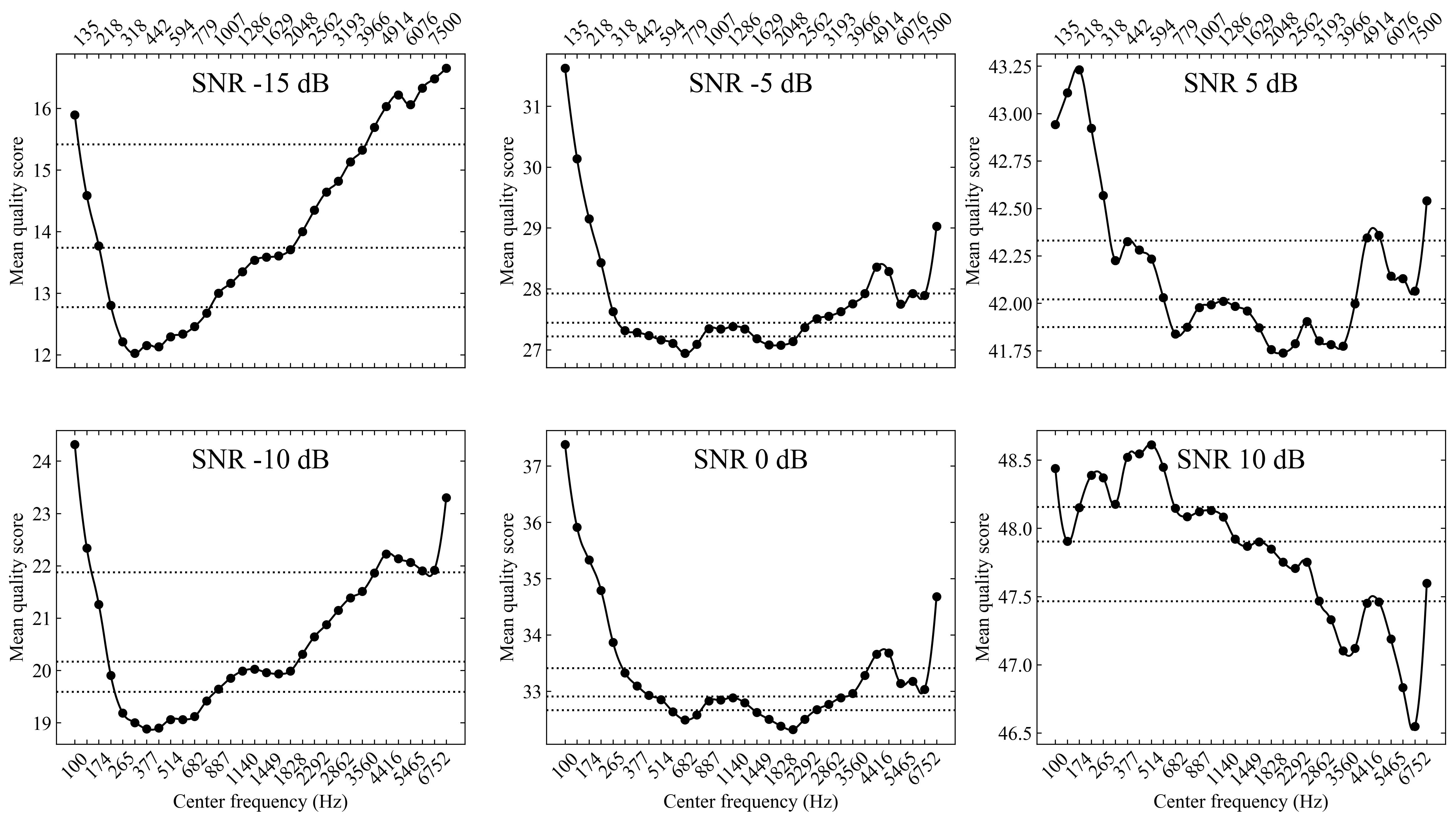}
\caption{Mean perceptual speech quality scores as a function of the center frequencies of the 32 bands at all 6 SNR conditions in the preliminary experiment. Scores are calculated based on the approach introduced in Section~\ref{subsec:2:4}. The circle markers denote the actual scores. The smoothed curve is based on the B-spline interpolation of the actual scores with a degree of 2. The dotted lines from the bottom to the top each represent the 25th, 50th (median), and 75th percentile of the 32 scores, respectively.}
\label{preliminary}
\end{figure*}

In the preliminary experiment, 51 unique speech materials were chosen to be tested at 6 different SNR values and 17 band combinations. Fig.~\ref{overall} presents the mean perceptual quality scores as a function of the SNR of the target bands averaged across all 17 different band combinations based on all accepted responses. Actual scores are represented by the circle markers. A second-order regression line denoted as the dashed line is fitted to indicate the overall pattern. Boxplots are also shown to visualize the general distributions of results in each SNR condition, by presenting the 25th, 50th (median), and 75th percentiles. Fig.~\ref{overall} suggests that the listening test was properly designed with reasonable SNR values and band combination size chosen so that sufficient difference could be perceived during the experiment with the lowest and highest mean quality scores shown by the anchors and the clean references, respectively. All other scores increase proportionally to the target bands' SNR values. This indicates that the subjects were assessing the qualities of the noisy speech signals at different SNR conditions as expected even with the implementation of the anchors modified and the labeled references absent, as it can be seen that the anchors were rated slightly lower than the -15 dB group and the clean references were rated much higher than the 10 dB group. It also suggests the data cleaning was properly done to eliminate undesired responses. Variances of the responses at each SNR suggested by the boxplots are gradually increasing, suggesting that the responses are more spread out with more uncertainty and that subjects rated more inconsistently as the SNR values increase. However, the anchor and reference groups seem to be two exceptions with them being rated more consistently. The group means and medians are closer to each other and the boxes also appear to be more symmetrical as the SNR values increase, which potentially suggests the responses are becoming more normally distributed. The clean reference group appears to be the exception to this pattern. All the group means, medians, 25th, and 75th percentiles increase proportionally to the SNR, which further strengthens the belief that the results are fundamentally reasonable and scientific.

\subsection{Preliminary experiment: perceptual quality scores by band and SNR}

All collected responses were grouped by different SNR conditions and band combinations. 
The band level quality scores were subsequently calculated following the approach introduced in Section~\ref{subsec:2:4}. Fig.~\ref{preliminary} presents the mean perceptual quality scores of 32 individual bands from band 0 to band 31 at all 6 SNR conditions. Actual scores are denoted by the circle markers on the curve smoothed by applying B-spline interpolation to the actual scores with a degree of 2. The dotted lines from the bottom to the top each indicate the 25th, 50th (median), and 75th percentile of the 32 quality scores at this SNR condition, respectively. For the -15 dB SNR condition, the overall quality scores appear to be highly non-uniform while a clear overall trend can be seen that the scores decrease first and increase later as the band center frequency increases. The lowest score is located in the band with a center frequency of 318 Hz. At -10 dB, the trend is highly similar to that in the previous one. The quality scores in this group are overall higher as the overall quality of the audio clips in this group appears to be better. A similar overall trend can be observed with the lowest score located in the neighboring band of the previous one with a center frequency of 377 Hz. Similar to what can be observed between the -15 and -10 dB conditions, the trends shown in the -5 and 0 dB groups resemble each other well. In contrast to the previous two conditions, a large portion of the bands in the mid-frequency region show similar quality scores lower than those in the low and high-frequency regions. It also appears that the low-frequency region in these two conditions has much higher quality rating scores than the other regions. The scores at the 5 and 10 dB conditions suggest different trends in that the scores appear much more irregular both in small areas and across the whole spectrum. An especially unique trend unlike others can be observed in the 10 dB group where overall the quality scores decrease as the center frequencies increase. The differences between the highest and the lowest quality scores are 4.63, 5.43, 4.69, 5.06, 1.49, and 2.06 from -15 to 10 dB SNR conditions, respectively. It is obvious the scores in 5 and 10 dB are smaller than the other groups. Fig.~\ref{overall} also shows the variances in these two groups are larger. It is therefore suspected that the results obtained from these two groups are too noisy to be considered significant. Overall, results from all six groups at different SNR conditions suggest certain degrees of similarities and differences. The mid-frequency region in all five groups except the one at 10 dB appears to be more likely to be associated with lower quality scores. All six groups present inconsistencies with fluctuating quality scores in small frequency regions and across the spectrum.

\subsection{Primary experiment: perceptual quality scores by SNR}

\begin{figure}
\centering
\includegraphics[width=0.75\columnwidth]{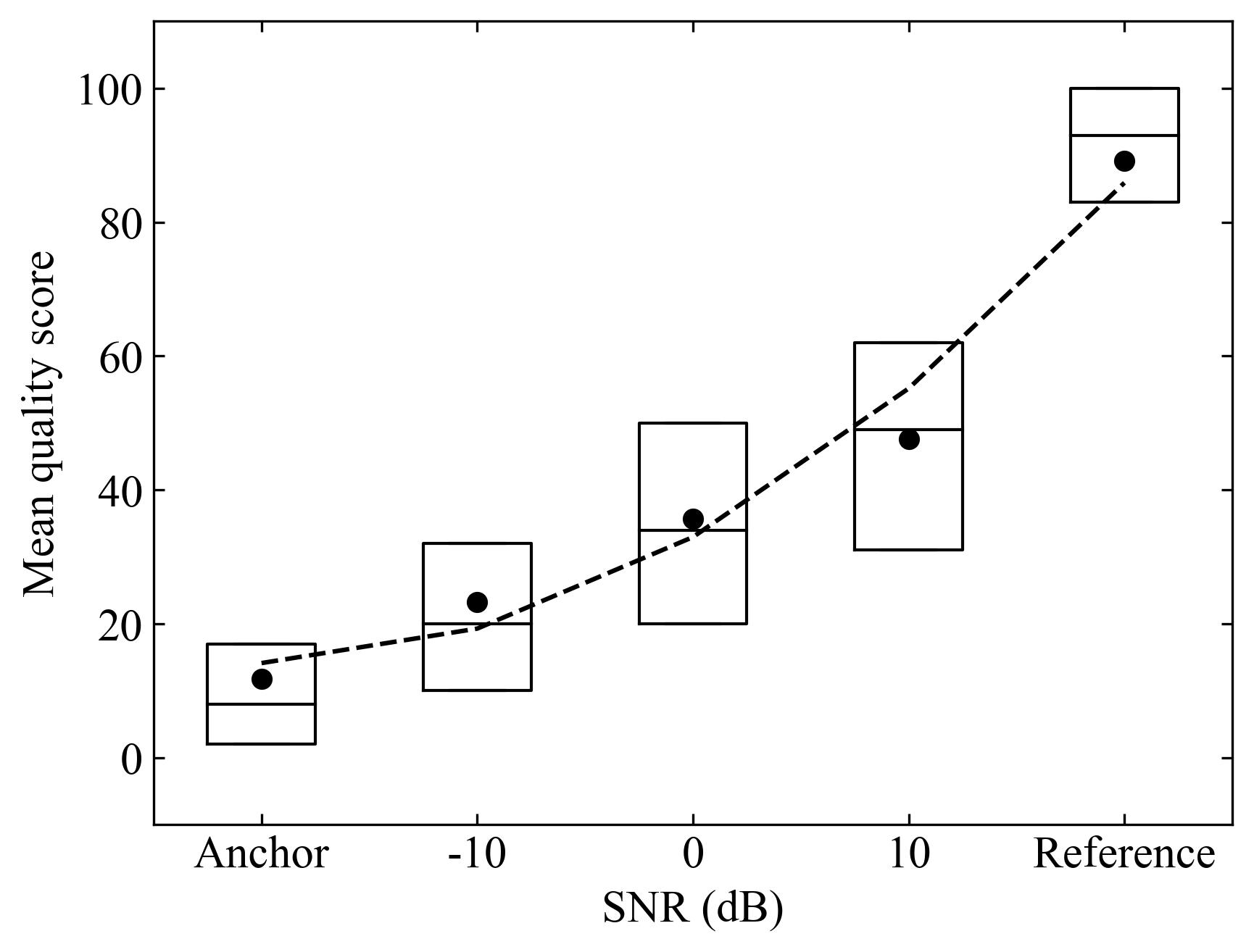}
\caption{Mean perceptual speech quality scores (e.g., circle markers) based on the listening test as a function of the SNR of the target bands in the primary experiment. Scores are calculated by averaging across all 32 band combinations.}
\label{overall_c16}
\end{figure}

The primary experiment was conducted to ensure that the low and high-frequency bands were evaluated as often as the mid-frequency bands. This, however, causes the target bands to be disconnected in some cases and is less ideal to reflect real-world situations. The number of audio clips from each noise category was also reduced and balanced. Based on the observation from the preliminary experiment, only 3 SNR conditions were investigated due to the high levels of similarities between groups of the results. In this experiment, 5 speech materials were chosen from each noise category with a total number of 20 materials being tested at 3 different SNR conditions and 32 different band combinations. Fig.~\ref{overall_c16} presents the mean perceptual quality scores as a function of the SNR. The results highly agree with those in the preliminary experiment in its values, overall trends, etc. However, it occurs that the quality scores at 10 dB become more skewed compared to the previous experiment, which suggests a higher level of asymmetry in its score distribution.

\begin{figure*}
\centering
\includegraphics[width=\textwidth]{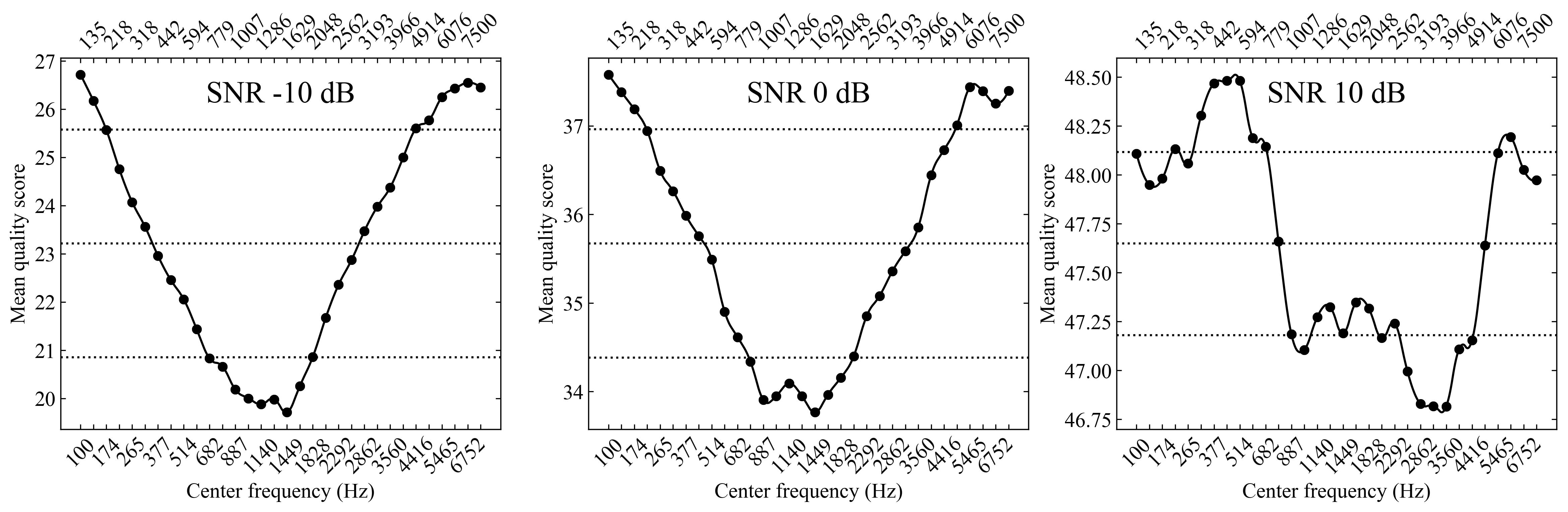}
\caption{Mean perceptual speech quality scores of the 32 bands at all 3 SNR conditions in the primary experiment.}
\label{primary}
\end{figure*}

\subsection{Primary experiment: perceptual quality scores by band and SNR}

Fig.~\ref{primary} shows the mean perceptual quality scores at 3 SNR conditions. At -10 and 0 dB, the overall trends appear to be more regular compared to the previous experiment, where the scores in the mid-frequency region are, similar to some of the previous observations, lower than those in other regions and the plots overall appear to be smoother with fewer fluctuations in small areas. The trends given by the two groups are also more similar to each other. Moreover, the differences in the mean quality scores of the first and the last few bands become much smaller due to the circular operation in combining the target bands. This can be especially seen from the differences in the results at the 0 dB conditions of the two experiments. It should be noted that, however, the regions with the lowest quality scores now appear in the higher frequency region with the lowest scores observed at the band with a center frequency of 1449 Hz in both the -10 and 0 dB conditions. Results from the 10 dB group again appear to be slightly different with more obvious fluctuations and the lowest scores are observed in regions with even higher center frequencies around 3560 Hz. However, it can be seen that it agrees slightly better with the other two SNR groups in terms of the overall trend across the spectrum.

\subsection{Perceptual quality index scores by band}

\begin{figure*}
\centering
\includegraphics[width=0.70\textwidth]{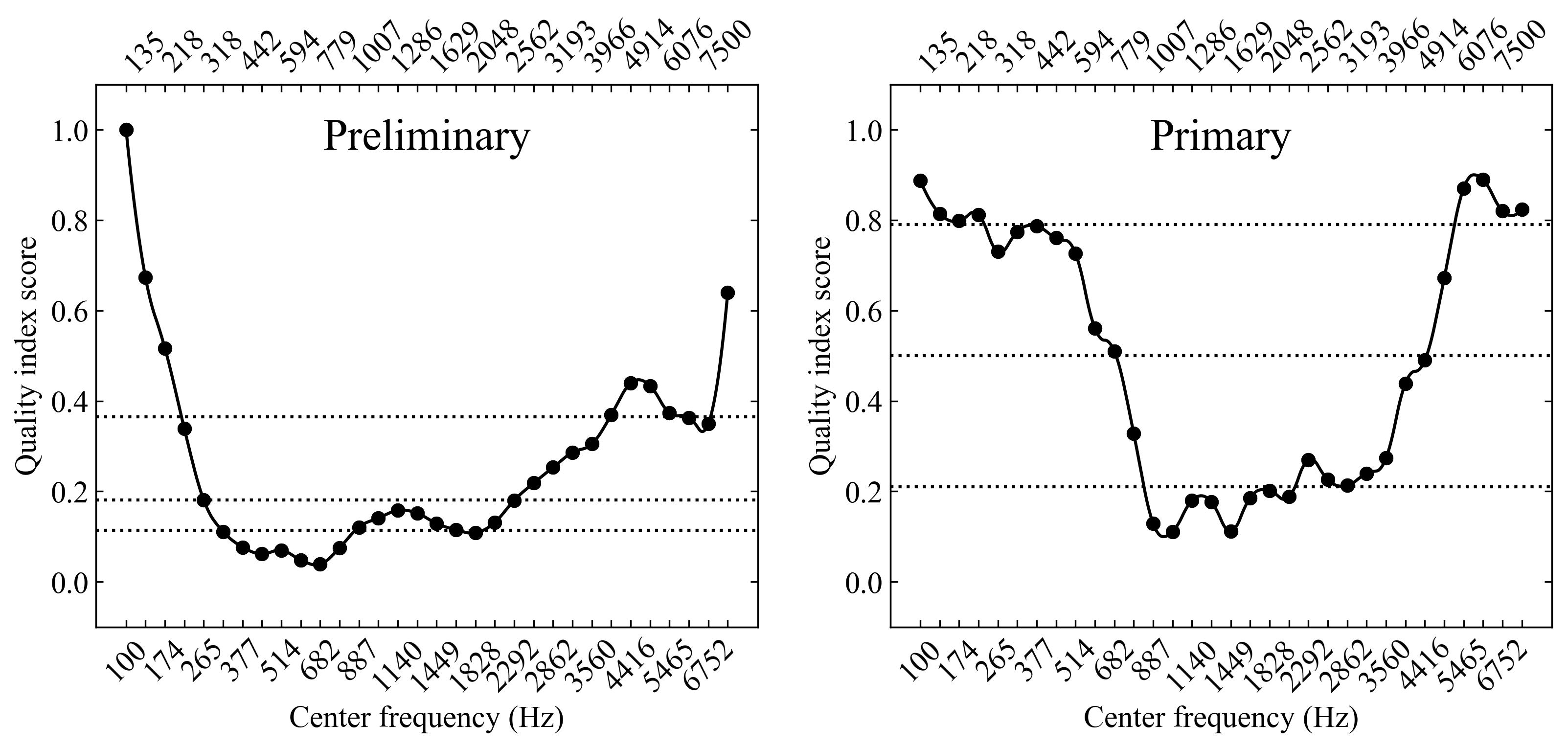}
\caption{Quality index scores averaged across the selected SNR conditions in the preliminary and primary experiments.}
\label{indexfigure}
\end{figure*}

To further generalize the conclusions and incorporate all essential groups of results above, perceptual quality index scores by target band were calculated according to Section~\ref{subsec:2:4}. Only conditions at SNRs of -10 and 0 dB were used for the calculation, without considering those groups that are either highly similar to the two chosen or not significant enough to be representative. All the original mean perceptual quality scores were scaled to values between 0 to 1 by min-max normalization. The index score of a certain band was calculated by averaging across all three normalized quality scores of the same band at the three chosen conditions. Plotted in a similar fashion to the previous ones, Fig.~\ref{indexfigure} presents the quality index scores for all the 32 target bands from the two experiments. It is believed that the index scores are a summary and an average of all quality scores in the three chosen SNR conditions. One major difference between the two groups of results is that the quality scores of the first and the last band in the primary experiment are much closer to each other compared to those in the preliminary experiment. This is again very likely due to the circular operation employed when combining the target bands. If the same operation was applied to the results in the preliminary experiment, the results in the two groups would appear more similar by both presenting higher quality scores in the low and high-frequency regions and lower quality scores in the mid-frequency region. Fluctuations can still be observed in small regions across the spectrum in both groups.

\section{Discussion}
\label{sec:4}

The results in the previous section provide the perceptual quality scores obtained from the properly designed experiments. As the quality scores obtained in this way have been shown to be highly correlated to the true perceptual speech quality, it is believed that the results are capable of reflecting the robustness of individual frequency bands to noise to a certain extent.

Figures~\ref{overall} and \ref{overall_c16} present the statistical summaries of the median, 25th, and 75th percentiles at each individual noise level. The general pattern of how subjects tended to rate the target signals in the experiments is revealed by these values. In fact, the patterns and these statistical summaries based on the individual results at all 17 or 32 band combination conditions are believed to follow the same trend, validated by Pearson's correlation which gives statistically significant high correlation coefficients between any two combination conditions. It is noticed that the variance of the examined groups at each SNR condition in general increases as the SNR itself increases, indicated by the distance between the 25th and 75th percentiles. Exceptions appear at the anchor and reference conditions because they were not processed at the band level. This observation potentially suggests the audio signals from groups at higher SNRs (i.e. 5 and 10 dB) may sometimes be too clean for the current study and therefore the participants found it challenging to rate these signals. This assumption is later suggested again by results at 5 and 10 dB conditions in Figures~\ref{preliminary} and \ref{primary}, with those groups of results showing different patterns. 

To further investigate the significance of results at all SNR conditions, for each signal rated in the listening study, one-half of its responses were randomly selected to generate plots shown in Figures~\ref{preliminary} and \ref{primary}, which are then compared to the plots generated based on the other half of the responses. The results shown by Pearson's correlation supported the assumptions by rendering only $\rho=0.13$ and $\rho=0.35$ for the SNR 5 and 10 dB conditions in the preliminary experiment, respectively, with all $\rho$ for other SNR groups in both experiments being larger than 0.70. This potentially suggests those two SNR conditions in the preliminary experiment failed to provide statistically significant results, possibly due to the underlying assumption that the quality differences of the audio signals in the positive SNR groups are too hard to perceive as they become too clean. This assumption was further confirmed by comparing the average widths of the 90\% confidence intervals of the quality scores at 32 center frequencies at a certain SNR group to the differences between the highest and the lowest quality scores within that specific group. For SNR 5 and 10 dB in Figures~\ref{preliminary} and SNR 0 and 10 dB in Figures~\ref{primary}, their individual average confidence intervals are larger than their differences between the highest and lowest scores. Given results from both significance tests, it is determined that the results from SNR 5 and 10 dB in Figures~\ref{preliminary} and 10 dB in Figures~\ref{primary} are too noisy to be considered significant. They were not included in any of the quality index score calculations. Particularly, among the different individual SNR conditions in both experiments, similarities and differences can be observed. Examined by Pearson's correlation, results from groups at -15 dB and -10 dB in the preliminary experiment are highly correlated with $\rho=0.92$. Results from groups at -5 dB and 0 dB are also correlated with $\rho=0.96$. A similar conclusion was also observed between -10 dB and 0 dB in the primary experiment ($\rho=0.97$). Such observations also helped reduce the SNR conditions that needed to be examined in the primary experiment. This, however, suggests that the robustness of frequency bands may behave differently under the influence of different levels of noise. ANOVA tests were used to compare the mean of the quality scores from different SNR groups. For both the preliminary and primary experiments, the ANOVA tests provided p\textless0.05. It can be therefore concluded that there are significant differences among different SNR groups. Tukey's Honestly Significant Difference (HSD) test also confirms that such differences can be observed in any pairs of two different SNR groups in the preliminary or the primary experiments, with p\textless0.05 in all cases.

The main difference between the preliminary and the primary experiment is the circular operation in creating the target band groups. The operation allows each individual band to be examined the same amount of times in the experiment. However, it is unclear whether the results obtained from these audio signals are necessarily a better reflection of the noise robustness of the frequency band in real situations where frequency bands in the low and high-frequency areas are not necessarily closely associated with each other. Due to this operation, only a moderate correlation between the quality index scores of the preliminary and primary experiments can be observed as suggested by Pearson's correlation with $\rho=0.49$. Overall, from what can be observed in Figures~\ref{preliminary} and \ref{primary}, frequency bands in the mid-frequency region appear to be less robust to noise compared to those in the low and high-frequency regions. In general, the overall quality scores increase as the amount of distortion decreases (i.e.~as SNR increases). These results are consistent with the findings from \citet{hansenoriginal}, where it shows that the least distorted conditions are ranked higher at each center frequency. However, the preference quality ratings do not vary much across center frequencies, which differs from the current findings. This may occur since more center frequencies were evaluated in the current study, thus allowing more granular responses, and since the participants were asked to provide quality scores but not perform pairwise preferences. Also, real-world noises were considered (not modulated white noise), with varying types of sounds that have different frequency responses.

Given the discussions above, it is believed that under different levels of compromising noise, the band-level noise robustness can behave differently to a certain extent and different strategies should be deployed accordingly when tackling noise in the hope of improving speech quality. However, Fig.~\ref{indexfigure} can still provide meaningful information in general when the noise level is unknown or the noise level covers a wide range of SNRs. Similar to what was observed in \citet{sarah}, the low-frequency region appears to be most robust to noise in both experiments. However, the mid-frequency region is observed to be the one least robust to noise in general with the high-frequency region providing mediocre noise robustness in the preliminary experiment or a similar noise robustness to the low-frequency region in the primary experiment. Fluctuations can be observed in results from both current experiments, although they appear to be less obvious compared to those shown in the previous intelligibility study, which may potentially result from the inherent difference in speech quality and intelligibility. The results from the two studies again suggest the intricate relationships between these two metrics by presenting both similar and different observations. Despite the noticeable overall trends, it can be concluded based on the results that the pattern of noise robustness is not simple as shown across the spectrum where inconsistency can be spotted throughout the frequency bands.

The quality index scores were also investigated by noise categories to study if noise types may potentially have different impacts on the noise robustness of the frequency bands. Given the fact that these categories of noise may present drastically different audio features, it is suspected that they compromise the band-level speech quality in different ways which leads to different noise robustness behaviors. Similarities and differences can both be observed among results from the four noise categories in the primary experiment. A strong correlation between STR and CAF can be observed and examined by Pearson's correlation with $\rho=0.78$. Strong correlations between CAF and BUS were also suggested by Pearson's correlation with $\rho=0.88$. Interestingly, a strong negative correlation was also spotted between PED and BUS with $\rho=-0.77$. ANOVA test results also confirm that there are statistically significant differences among the means of the four groups of results obtained from different noise types, with p\textless0.05. The results altogether suggest that different types of noise can affect band-level noise robustness differently. This may result from the distinct audio features that different types of noise possess, or from some other high-dimensional intricate details in noise and how they interact with the human auditory system. It is unclear what exactly causes the different robustness behaviors, but the results suggest the experiments on which the conclusions are based should carefully consider comprehensive categories of noise.

\begin{figure}
\centering
\includegraphics[width=0.75\columnwidth]{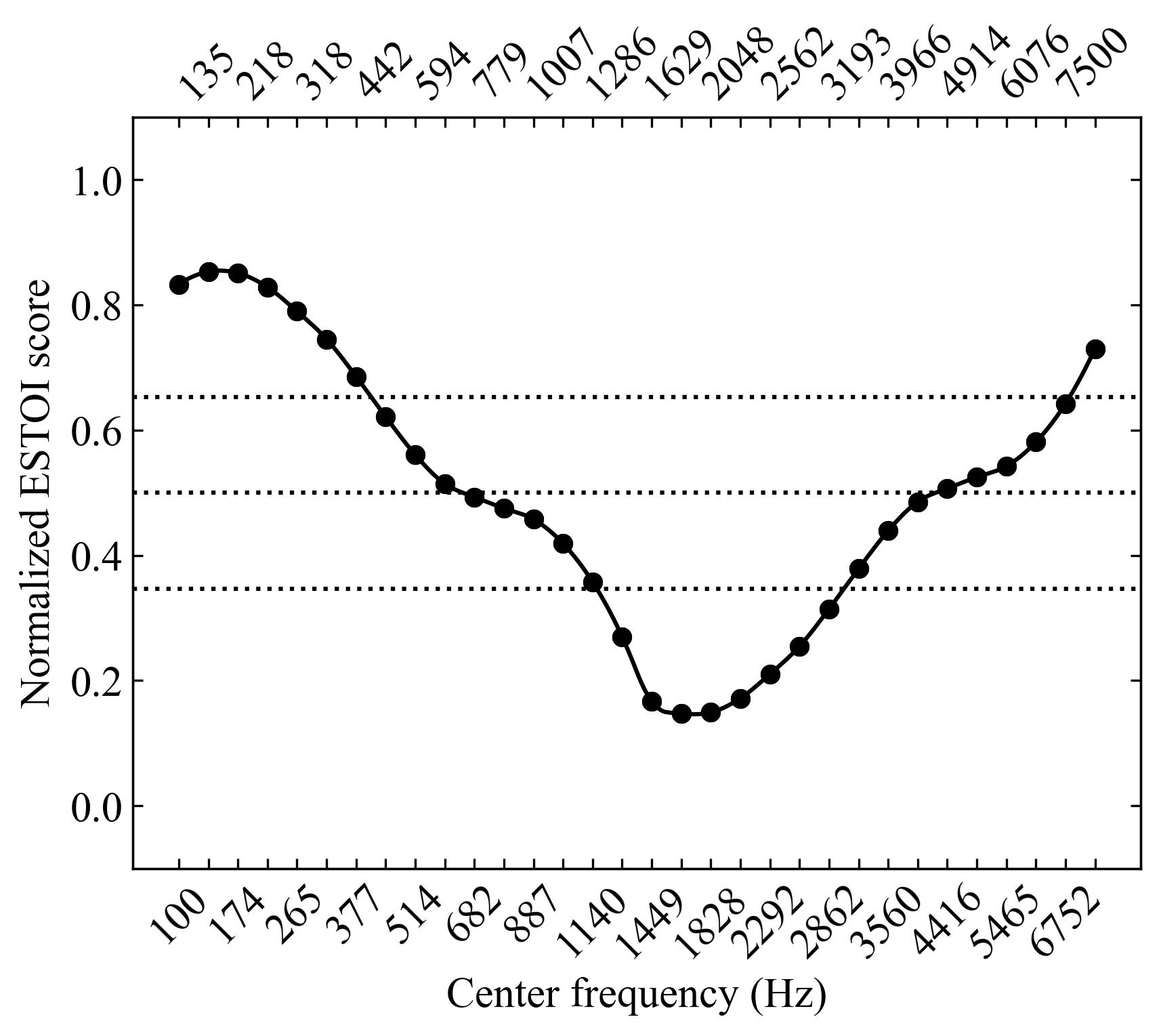}
\caption{Normalized ESTOI scores of the target bands.}
\label{indexestoi}
\end{figure}

More experiments were further deployed in the hope of investigating how well the popular objective metrics including PESQ \cite{pesq}, STOI \cite{stoipaper}, and ESTOI \cite{estoipaper} perform, where the perceptual quality differences among the test signals can be exceedingly subtle. Results were based on the audio signals used in the primary experiment and processed based on the same approach introduced in Section~\ref{subsec:2:4} whereas the only difference is that the mean subjective quality rating score for each test audio signal was replaced by the objective quality score. Compared to the results in the primary experiment, only ESTOI was capable of generating similar patterns in all three SNR conditions as can be examined by Pearson's correlation. Fig.~\ref{indexestoi} presents the normalized ESTOI scores calculated from the three SNR conditions. A correlation between normalized ESTOI and the primary experiment scores can be observed and examined by Pearson's correlation with $\rho=0.87$, while PESQ was only given $\rho=0.45$ and and STOI was given $\rho=0.20$, with their figures not shown for brevity. However, the trends suggested by the scores given by ESTOI and the primary experiment do not highly resemble each other, with the ESTOI suggesting a narrower mid-frequency region where most of the low-quality index scores occur. This can be partially because ESTOI is a metric used to evaluate speech intelligibility, whereas the results in the current study are based on speech quality, which is a higher-dimensional and more complicated feature and therefore can be affected by more factors. Overall, it is suspected that the current objective metrics commonly deployed in various situations to assess speech quality and intelligibility are not fully capable of providing significant results on difficult tasks such as the one in this study. Subjective listening tests still have their superiority on these occasions. More advanced objective speech quality assessment metrics that produce results better correlated to those obtained from subjective methods are desired. 

The current approach has successfully uncovered the subtle differences in the noise robustness of various frequency bands. Previous research has shown that MUSHRA can detect minute differences in speech quality by collecting human responses in various properly designed listening tests. With the appropriate configuration of the speech band combination and the proper selection of target band SNRs, differences among the band robustness were revealed with the support of a large number of quality scores. However, the listening tests nevertheless still required a substantial amount of subjects and time, as the band-level examination considerably increased the scale of the experiments. It is hoped that a more efficient methodology can be proposed in the future to study this topic with only fewer participants and less time required while still providing significant results. 

The findings in the current study will contribute to the understanding of how noise may impact speech quality at the frequency band level. The relationships between speech quality and intelligibility have been discussed frequently where similarities and differences can be observed from various perspectives. It is hoped that in future research the way noise impacts speech quality and intelligibility can be better associated and summarized based on the current and past studies and that simple solutions to how to tackle noise from the perspective of both quality and intelligibility can be provided, as improving both of them is considered to be significant to many research topics. For instance, the design of deep learning models deployed in areas such as speech enhancement can incorporate the conclusions in this study where certain frequency regions are believed to be more vulnerable to noise. This may benefit these research studies by providing more efficient but more powerful models that require less computational capacity but render better outcomes at the same time. It is also hoped more efforts in studying speech quality can be made in future research, which can be particularly beneficial to those including the telecommunication industry and the community for individuals with hearing-impairments.

\section{Conclusions}
\label{sec:5}

We investigated the robustness of frequency bands to noise based on speech quality. Perceptual listening tests inspired by MUSHRA were deployed to assess the speech quality of broadband real-world speech signals compromised at the frequency band level by real-world noise at different SNR conditions. The findings are as follows.

1. The robustness of frequency bands to noise was observed to be non-constant across the spectrum.

2. The overall pattern of noise robustness and how it impacts speech quality does not have a simple answer, although general trends can be concluded that the low and high-frequency regions appear to be more robust to noise and the mid-frequency region appears to be less robust. Fluctuations are observed across the spectrum at various SNRs.

3. Relationships of how noise impacts quality and intelligibility can be observed, although no strong correlations were spotted and major differences exist.

4. Different categories of noise impact the speech quality differently. General conclusions of noise robustness of frequency bands should come from experiments based on as comprehensive categories of noise as possible. Otherwise, the topic of noise robustness should be specified and restricted to ``what type'' of noise robustness.

5. Some current objective quality and intelligibility metrics do not provide statistically significant results on difficult tasks. More advanced objective metrics are needed. 

6. The deployed listening test reveals the minute differences in band-level noise robustness, with appropriate setups of the groupings of frequency bands and the proper selection of SNR values.

In the future, how the findings in this study will instruct and benefit the techniques in speech enhancement will be investigated. It is hoped that the concept can be incorporated into the design of future speech enhancement techniques to both reduce the scale of the models and fine-tune the performance.

\section*{Acknowledgments}

This work was supported by NSF under Grant IIS-1942718. We also thank the supercomputing cluster provided by Indiana University for data processing.

\section*{Author Declarations}

\subsection*{Conflict of Interest}
The authors do not have conflicts to disclose.

\subsection*{Ethics Approval}
Ethics Approval was obtained from the Ohio State University's Office of Responsible Research Practices. Informed consent was obtained from all participants.

\subsection*{Data Availability}
The authors are unable to redistribute the stimuli used in the study due to copyright restrictions.

\bibliographystyle{plainnat}
\bibliography{junyijasa}

\end{document}